\documentclass[9 pt,twocolumn]{revtex4}

\usepackage[usenames, dvipsnames]{color}
\usepackage{graphicx,epstopdf}
\usepackage{leftidx}
\usepackage{amsmath}
\usepackage{tensor}
\usepackage{graphicx}
\usepackage{dcolumn}
\usepackage{bm}
\usepackage{hyperref}
\usepackage{amssymb}
\usepackage{hyperref}
\RequirePackage[normalem]{ulem} 
\RequirePackage{color}\definecolor{RED}{rgb}{1,0,0}\definecolor{BLUE}{rgb}{0,0,1} 
\immediate\write18{texcount -char -freq
 borra.tex > /tmp/charcount.tex}

\begin{document}
\newcommand{\be}{\begin{equation}}
\newcommand{\ee}{\end{equation}}
\newcommand{\ba}{\begin{eqnarray}}
\newcommand{\ea}{\end{eqnarray}}
\newcommand{\Gam}{\Gamma[\varphi]}
\newcommand{\Gamm}{\Gamma[\varphi,\Theta]}
\thispagestyle{empty}

\title{  Bipartite correlations in quantum resonance states  }

\author{Przemys\l aw Ko\'{s}cik }

\address{Institute of Physics,  Jan Kochanowski University,
ul. \'Swi\c{e}tokrzyska 15, 25-406 Kielce, Poland}

\begin{abstract}
We discuss   a  diagonal representation of a    reduced density matrix  determined within the framework of the complex scaling
method. We also discuss a possible measure  of  bipartite correlations
in quantum resonance states. As an  example, we consider a one-dimensional  system of two  bosons with a contact interaction
  subjected to  an open  potential well. The correlation properties of the  lowest-energy resonance state of the system are explored over a wide range of  the inter-boson interaction strength, including the   Tonks-Girardeau regime.

\end{abstract}

\maketitle

\section{Introduction}

Quantum entanglement  is a  fundamental
feature of the quantum world and has attracted significant  research attention. New experimental techniques  have
opened up opportunities for practical applications  of quantum entanglement  in various branches of information technology \cite{p,p2}.  The fields of  quantum teleportation, quantum cryptography and quantum computation have grown particularly rapidly    in the last few years.
Entanglement    is  also  used as an alternative measure of  correlation in   systems of interacting particles \cite{cor}.
Considerable efforts have been made to understand quantum correlations  in bound states of   model systems such as the Moshinsky atom \cite{mon0,mon1,mon2,mon3,mon4}, quantum dot systems \cite{ent1,kos3,kos4,kos5,kos6} or    ultra-cold boson systems \cite{bos0,bos1,bos2,bos22,bos3}. Moreover, in  recent
years,   the helium atom and helium-like ions have been  extensively studied in this context
\cite{hel,helkkk,hel0,hel1,hel2,hel3,hel4,hel5,hel6,hel8,hel89}. For an overview of the
recent developments in studies of  entanglement in quantum composite
systems,  see \cite{tich}.

However,
relatively few attempts have been made    to improve
understanding of the correlation properties of  systems that exhibit
metastable states. Various theoretical methods to determine  resonance energies and lifetimes can be found in the literature. Among
these, the most popular are the    complex scaling method (CSM)
\cite{1} and the real stabilization method \cite{ha}.
Treatment of the entanglement  of resonance states with the CSM     was  proposed in \cite{ent1},  where   both a complex-scaled density operator and a complex linear entropy  were introduced. Within the framework of this formalism, the  resonance states of   two-electron    Gaussian quantum dots
   have recently been  analysed from the perspective of quantum  information \cite{kuros}.
Nonetheless, as far as we know,  no study  has discussed in detail the  diagonal representation   of the complex-scaled reduced density matrix. This gap in the literature provides the motivation for the present Letter.

The remainder of this Letter is structured as follows.  Section \ref{sec1}  briefly outlines   the CSM.  Section  \ref{jjl}   discusses  the   diagonal form of the   reduced density matrix determined in the framework of the CSM.  Section \ref{entg} discusses  possible  correlation measures for    resonance states.  Section \ref{ex} focuses on  the correlation properties of   systems that contain  two interacting bosons trapped inside an open potential well. Finally, Section \ref{con} presents concluding remarks.

\section{The complex scaling method}\label{sec1}

The CSM is a powerful tool  for searching for the resonant parameters of a system that supports   metastable states \cite{1}. The  utility
of this method  is that the resonant
states  can be treated  by applying the methods used to compute
bound states, for example,   a finite-basis-set approximation  \cite{kk,kk1}.
Here we only  briefly outline the complex scaling formalism. Details of the computational technique of the CSM can be found in an excellent overview  \cite{1}.

In the CSM, the   Hamiltonian $\hat{H}= \hat{T}(\textbf{x} )+\hat{V}(\textbf{x} )$, where $\hat{T}$ and $\hat{V}$ are  kinetic and potential energy operators, respectively,  is  transformed
by  coordinate transformation
$\textbf{x}\mapsto\textbf{x} e^{I \theta}$ into \be \hat{H}^{\theta}=e^{-2 I \theta} \hat{T}(\textbf{x})+\hat{V}(\textbf{x} e^{I \theta}),\label{comh} \ee
where $\theta$ is the so-called scaling angle. A key aspect of the CSM is that the complex-scaled   Hamiltonian $\hat{H}^{\theta}$ is a non-Hermitian operator
and the inner product is defined as
\be \ll\psi|\varphi\gg=\int_{\textrm{all space}} \psi(\textbf{x})\varphi(\textbf{x})d\textbf{x},\label{ii}\ee
$\ll \textbf{x}|\varphi\gg=\varphi(\textbf{x})$ and $\ll\psi|\textbf{x}\gg=\psi(\textbf{x})$. The right ($R$) and left ($L$)  eigenstates of  $\hat{{H}}^{\theta}$ are defined by  $ \hat{H}^{\theta}|\psi^{R}_{i} \gg=W_{i}|\psi^{R}_{i} \gg$ and $\ll \psi^{L}_{i}| \hat{H}^{\theta}=\ll \psi^{L}_{i}|W_{i}^{'}$, respectively.
 These equations
   can be turned  into  algebraic problems by  expanding $|\psi^{R}_{i} \gg$ and $\ll \psi^{L}_{i}|$
as linear combinations of  $|\varphi_{k}\gg$ and $\ll\varphi_{k}^{*}|$,  respectively,  where $\{\varphi_{i}\}$ is a complete set of  orthonormal functions, $\ll\varphi_{i}^{*}|\varphi_{j}\gg=\delta_{ij}$.
Thus,
    the first equation becomes the  eigenvalue problem
    $(\textbf{H}-\textbf{I} W)\vec{R}=\vec{0}$ with $\textbf{H}=[{\ll \varphi_{i}^{*}|\hat{H}^\theta|\varphi_{k}\gg}]$, the eigenvalues of which $W_{diag}=diag(W_{0},W_{1},...,)$ (if
\textbf{H} is diagonalizable) are      given by $W_{diag}=\textbf{R}^{-1}\textbf{H}\textbf{R}$,
   where  $\textbf{R}$ is the  eigenvector matrix of  $\textbf{H}$. The second equation becomes the eigenvalue problem $({{\textbf{H}^{T}}}-\textbf{I} W^{'})\vec{L}=\vec{0}$. Because  $W_{diag}=W_{diag}^{T}=(\textbf{R}^{-1}{\textbf{H}}\textbf{R})^T=({\textbf{H}}\textbf{R})^T (\textbf{R}^{-1})^T=\textbf{R}^{T}{\textbf{H}^{T}}(\textbf{R}^{-1})^T$, it immediately follows that  the   eigenvector matrix of the matrix  $\textbf{H}^{T}$ is $\textbf{L}=(\textbf{R}^{-1})^{T}$ and $W_{i}=W_{i}^{'}$. As a result, we obtain
\be \ll \textbf{x}|\psi^{R}_{i}\gg=\psi^{R}_{i}(\textbf{x})=\sum_{k}(\textbf{R})_{ki}\varphi_{k}(\textbf{x}),\label{aa11}\ee
and \be  \ll\psi^{L}_{j}|\textbf{x}\gg=\psi^{L}_{j}(\textbf{x})=\sum_{k}(\textbf{R}^{-1})_{kj}^{T}\varphi_{k}^{*}(\textbf{x}).\label{bb11}\ee
 Because $\textbf{R}^{-1}\textbf{R}=\textbf{I}$, the family $\{\psi^{L}_{i},\psi^{R}_{i}\}$ forms a complete set  of orthonormal functions with respect to the inner product (\ref{ii}), that is, $\ll\psi_{i}^{L}|\psi_{j}^{R}\gg=\delta_{ij}$.
In particular, if $\{\varphi_{i}\}$ is a real basis and $\textbf{H}$ is a symmetric matrix ($\textbf{H}=\textbf{H}^{T}$), that is, $W_{diag}=W_{diag}^T=(\textbf{R}^{T} \textbf{H}\textbf{R})^T=\textbf{R}^{T} \textbf{H}^T\textbf{R}$  ($\textbf{R}^{-1}=\textbf{R}^{T})$,     then the right and left  wavefunctions can   alternatively be expressed in the same form, namely
$ \psi^{L,R}_{i}(\textbf{x})=\sum_{k}{(\textbf{{R}})_{ki}}\varphi_{i}
(\textbf{x})$, where $\textbf{R}^{T}\textbf{R}=\textbf{I}$.
 However, in the case of $\textbf{H}^T=\textbf{H}^{*}$ ($\textbf{H}$ is  Hermitian), that is, $W_{diag}=W_{diag}^*=(\textbf{R}^{\dag} \textbf{H}\textbf{R})^{*}=\textbf{R}^{T} \textbf{H}^*\textbf{R}^*=\textbf{R}^{T} \textbf{H}^T\textbf{R}^*$ ($\textbf{R}^{-1}=\textbf{R}^{\dag}$),
 they can be written as $\psi^{R}_{i}(\textbf{x})=\sum_{k}(\textbf{R})_{ki}\varphi_{k}(\textbf{x})$ and
$ \psi^{L}_{i}(\textbf{x})=\sum_{k}{(\textbf{R}^{*})_{ki}}\varphi_{k}^{*}(\textbf{x})$, where $\textbf{R}^{\dag}\textbf{R}=\textbf{I}$. Note that our conclusions
coincide with those of \cite{1}.

The resonances  appear as  $\theta$-independent  complex eigenvalues $W_{k}$ with $\mbox{Im}[W_{k}]<0$  \cite{ABC}, and the resonance energies $E_{k}^{rez}$ and  lifetimes $\Gamma_{k}$ are obtained as $E_{k}^{rez}=\mbox{Re}[W_{k}]$ and $\Gamma_{k}=-2\mbox{Im}[W_{k}]$, respectively: \be W_{k}=E_{k}^{rez}-I {\Gamma_{k}\over 2}.\ee

\section{Complex-scaled reduced density matrices}\label{jjl}

Suppose we divide a system  into two  parts, $A$ and $B$. Let us express
the right and left wavefunctions of a given resonance
state as follows:
\be |\psi^{R}\gg=\sum_{ij}e_{ij}^{R}|a_{i}\gg_{A}|b_{j}\gg_{B},\label{op0}\ee
and \be \ll\psi^{L}|=\sum_{kl}e_{kl}^{L}\tensor*[_A]\ll{} a_{k}^*|\tensor*[_B]\ll {}b_{l}^*|.\label{op1}\ee
Here, ${\ll \psi^{L}|\psi^{R}\gg}{}=1$, where $\{a_{j}\}$ and $\{b_{j}\}$  are bases of square integrable  orthonormal functions for
 the subsystems $A$ and $B$, respectively, $_{A}{\ll {}a_{i}^{*}|a_{j}\gg}{_A}=\delta_{ij}$, $_{B}{\ll b_{i}^{*}|b_{j}\gg}{_B}=\delta_{ij}$.
Following  \cite{ent1}, we define the density operator as \be\hat{\rho}^{AB}=|\psi^{R}\gg\ll\psi^{L}|.\ee
The reduced density matrix  of  subsystem  $A$ is thus obtained by tracing
out the $B$ degrees of freedom, which gives
\be \hat{{\rho}}_{A}=\mbox{tr}_{B}[\hat{\rho}^{AB}]=\sum_{ik} \rho_{ik}^{A}|a_{i}\gg_{AA}\tensor*[]\ll{} a_{k}^*|,\label{lopm}\ee
where $\rho_{ik}^{A}=\sum_{j} e_{ij}^{R}e_{kj}^{L}$. Its right and left eigenstates  are defined by $ {\hat{\rho}}_{A} |u_{n}^R\gg_{A}=\lambda_{n}|u_{n}^R\gg_{A}$
and $ \tensor*[_A]\ll{} u_{n}^L|{\hat{\rho}}_{A} =\tensor*[_A]\ll {} u_{n}^L|\lambda_{n}^{'}$, respectively.  In strict analogy with Section \ref{sec1}, we  conclude that
 \be|u_{n}^R\gg_{A}=\sum_{j}(\textbf{V})_{jn}|a_{j}\gg_{A},\label{jj1}\ee and  \be \tensor*[_A]{\ll u_{n}^L|}{}=\sum_{j}{(\textbf{V}^{-1})^{T}_{jn}}\tensor*[_A]{\ll a_{j}^*|}{}, \label{jj2}\ee
where $ \textbf{V}$ is the eigenvector matrix of the
 matrix $\rho_{A}=[_A\ll{} a_{i}^*|{\hat{\rho}}_{A}|a_{k}\gg_{A}]=[\rho_{ik}^{A}]$ with the eigenvalues $\lambda_{n}^{A}$ ($\lambda_{n}=\lambda_{n}^{'}=\lambda_{n}^A$),  $\tensor*[_A]\ll{} u_{n}^L|u_{m}^R\gg_{A}=\delta_{nm}$.
Following on from the above,   $\hat{{\rho}}_{A}$ can be  expressed in diagonal form as follows:
\be \hat{{\rho}}_{A}=\sum_{n}\lambda_{n}^{A}|u_{n}^R\gg_{AA}\ll u_{n}^L|,\label{a5a}\ee
  $[_A\ll{} u_{n}^L|\hat{\rho}_{A}|u_{m}^R\gg_{A}]=diag(\lambda_{0}^{A},\lambda_{1}^{A},...,)$. Because ${\ll \psi^{L}|\psi^{R}\gg}{}=1$, the normalization condition gives $\sum_{n}\lambda_{n}^{A}=1$ or equivalently $\mbox{tr} [{{\rho}}_{A}]=1$.
 Note  that  the eigenvalues of $\rho_{A}$ are generally  complex numbers,  except for the case $\rho_{A}=\rho_{A}^{\dag}$, that is,
when $\rho_{A}$ is a Hermitian matrix. Once again in strict analogy with Section \ref{sec1}, we can conclude that if  $\{a_{i}\}$ is a real   basis  and $\rho_{A}=\rho_{A}^{T}$, then the right and left eigenstates of $\hat{\rho}_{A}$  can  be expressed in an identical form, namely
  \be \tensor*[_A]\ll{} u_{n}^L|x\gg_{A}=_{A}\ll x|u_{n}^R\gg_{A}
=\sum_{k}{(\textbf{{V}})_{kn}}a_{k}(x),\ee $\textbf{V}^{T}\textbf{V}=\textbf{I}$.
Analogously, we  obtain the reduced density matrix for  subsystem $B$:
\be \hat{{\rho}}_{B}=\mbox{tr}_{A}[\hat{\rho}^{AB}]=\sum_{ik} \rho_{ik}^{B}|b_{i}\gg_{BB}\tensor*[]\ll{} b_{k}^*|,\label{lopm}\ee
 $\rho_{ik}^{B}=\sum_{j} e_{ji}^{R} e_{jk}^{L}$, and its diagonal form,
\be \hat{{\rho}}_{B}=\sum_{n}\lambda_{n}^B|v_{n}^R\gg_{BB}\ll v_{n}^L|,\label{a6a}\ee
with
 \be|v_{n}^R\gg_{B}=\sum_{j}(\textbf{U})_{jn}|b_{j}\label{jj3}\gg_{B},\ee and
 \be \tensor*[_B]{\ll v_{n}^L|}{}=\sum_{j}{(\textbf{U}^{-1})^{T}_{jn}}\tensor*[_B]{\ll b_{j}^*|}{}, \label{jj4}\ee
where  $\lambda_{n}^{B}$ are the eigenvalues of the
  matrix  $\rho_{B}=[_B{\ll{} b_{i}^*|{\hat{\rho}}_{B}|b_{k}\gg}{_B}]=[\rho_{ik}^{B}]$  and
   $\textbf{U}$  is the corresponding  eigenvector matrix, $\tensor*[_B]\ll{} v_{n}^L|v_{m}^R\gg_{B}=\delta_{nm}$.

 Let us rewrite the matrix  $\rho_{A}=[\rho_{ik}^{A}]$,
$\rho_{ik}^{A}=\sum_{j} e_{ij}^{R}e_{kj}^{L}$ and the matrix $\rho_{B}=[\rho_{ik}^{B}]$, $\rho_{ik}^{B}=\sum_{j} e_{ji}^{R} e_{jk}^{L}$
 as
$\rho_{A}=\textbf{e}^R(\textbf{e}^L)^T$ and $\rho_{B}=(\textbf{e}^R)^T\textbf{e}^L$, respectively, where $\textbf{e}^{L,R}=[e_{ij}^{L,R}]$. It is known that if $C$ and $D$ are square  complex matrices of the same size, then  the matrices $CD$ and $DC$  have the same eigenvalues \cite{hj}. We thus conclude that   $\rho_{A}$ and $\rho_{B}^T=(\textbf{e}^L)^T\textbf{e}^R$ have a common set   of eigenvalues. Hence,  bearing  in mind that  $\rho_{B}$ has the same eigenvalues as  $\rho_{B}^T$,  we arrive at the conclusion  that  the eigenvalues of $\rho_{A}$ and $\rho_{B}$ are identical, $\lambda_{i}^A=\lambda_{i}^B=\lambda_{i}$.

\section{Correlation measures}\label{entg}
  As discussed by Moiseyev \cite{1}, the real and imaginary parts of the mean value of
a given complex-scaled operator, $\ll \hat{Q} \gg\equiv\ll \psi_{L} |\hat{Q}|\psi_{R} \gg$, give the average value of the quantity under consideration,  and its  uncertainty, respectively.
It is easy to see that the average value of any operator acting on
 one of the subsystems, let it be  $A$,  is given by
\begin{eqnarray} \ll \hat{Q}_{A} \gg\equiv\ll \psi_{L}|\hat{Q}_{A}|\psi_{R} \gg=\sum_{ijk}e_{kj}^{L}e_{ij}^{R}\tensor*[_A]\ll{} a_{k}^* |\hat{Q}_{A}|a_{i}\gg_{A}.\label{lolo}\end{eqnarray}
Noting that $\sum_{j}e_{kj}^{L}e_{ij}^{R}=\rho^{A}_{ik}$, we obtain \be \ll \hat{Q}_{A} \gg =\sum_{ik}\rho^{A}_{ik}\tensor*[_A]\ll{} a_{k}^* |\hat{Q}_{A}|a_{i}\gg_{A}= \mbox{tr}[ \rho_{A} Q_{A}].\label{kk}\ee
Let us now address the question of how to characterize  the correlation in resonance states. Generally, according to the  standard quantum theory,  the von Neumann (vN) entropy $S=-\mbox{tr}[{\rho}_{A,B}\mbox{ln}{\rho}_{A,B}]$ \cite{von}  and the linear entropy  $S_{lin}=1-\mbox{tr}[{\rho}_{A,B}^2]$ \cite{linear}  are used to quantify the degree of entanglement  in  composite quantum systems. In a strict mathematical sense, the values of $S$ and $S_{lin}$ can be given by $S=-\langle \mbox{ln}\hat{\rho}_{A,B}\rangle$ and
$S_{lin}=\langle \hat{1}-\hat{\rho}_{A,B}\rangle=1-\langle \hat{\rho}_{A,B}\rangle$ \cite{fano}, where $\langle...\rangle$ is the conventional inner
product and $\hat{1}$ is the identity operator from the appropriate basis set. Accordingly, in the resonance case, we have \begin{eqnarray} S=-\ll {}\mbox{ln}{\hat{\rho}}_{A,B}\gg=-\mbox{tr}[{\rho}_{A,B}\mbox{ln}{\rho}_{A,B}]=\nonumber\\=-\sum_{i}\lambda_{i}\mbox{ln}\lambda_{i}, \label{entropy}\end{eqnarray} and
\begin{eqnarray} S_{lin}=1-\ll {}  {\hat{\rho}}_{A,B}\gg=1-\mbox{tr}[{\rho}_{A,B}^2]=\nonumber\\=1-\sum_{i}\lambda_{i}^2. \label{entropy1}\end{eqnarray}  Thus,   the    real and imaginary parts of
$S$, ($S_{lin}$) can be identified as  the  mean value of the  operator: $-\mbox{ln}\hat{\rho}_{A,B}, (\hat{1}-\hat{\rho}_{A,B})$  and its uncertainty, respectively.  In particular, the  real part of   $\ll \hat{\rho}_{A,B}\gg=\sum_{i}\lambda_{i}^2$ gives the average of the \emph{probability}  $\lambda_{i}$, whereas the imaginary part describes its uncertainty.
However, because the entropy  cannot rigorously be  treated  as the average value of a quantal observable \cite{entropyu}, the interpretation of  the  real part of $S$, ($S_{lin}$)  as the entanglement entropy of a resonance state is problematic.
 Despite of this fact, we call $S$ and $S_{lin}$   complex \emph{entropies}
  and propose them  as  measures of  \emph{correlation} between  the subsystems $A$ and $B$.    Thus, we   identify the real and imaginary parts of $S$, ($S_{lin}$) with the amount of    \emph{correlation} and the uncertainty of this amount, respectively.
  Note that the complex linear \emph{entropy} of  a resonance state was first  introduced
  in  \cite{ent1}.

 In a way that is analogous to the standard
quantum theory, the resonance state that is  factorized as a product of states can be regarded  as uncorrelated, this corresponds to the case in which only one eigenvalue is nonzero, $\lambda_{i}=1$, giving $S=S_{lin}=0$. Deviations from such a state can be characterized by the complex \emph{entropy}, as   discussed above.

\section{Example: two-boson system}\label{ex}
Thus far, we have kept our discussion quite general. As an  example, we
now consider  a simple model system composed of two identical bosons
interacting via a contact potential of strength $g$. For the sake of simplicity,  we model an external
potential  by $v(x)={0.5}x^2 e^{-{x^2/ 5}}$ so that the system does
not exhibit any bound state. The resonance parameters of this system can be found
from  the following  complex-scaled Hamiltonian  \cite{bos}: \be
{\hat{H}}^{\theta}=\sum_{i=1}^2[-{e^{-2 I\theta}\over
2}{\partial^2\over \partial
x_{i}^2}+v(x_{i}e^{I\theta})]+ge^{-I\theta}\delta
(x_{2}-x_{1})\label{ff}.\ee
Here we apply the basis function method and
diagonalize  the matrix
Hamiltonian
$\textbf{H}^{\theta}=\langle\phi_{nm}|\hat{H}^{\theta}|\phi_{ij}\rangle$
 in a basis of permanents constructed from the one-particle orthonormal basis set,
 \be \phi_{ij}=s_{ij}
[\psi_{i}({x}_{1})\psi_{j}({x}_{2})+\psi_{j}({x}_{1})\psi_{i}({x}_{2})], (i\geq j),\label{baza}\ee  where $s_{ii}\!=\!1/2$  and
$s_{ij}\!=\!2^{-1/2}$ for $i\neq j$. We choose
as the one-particle basis the wave functions of a simple harmonic oscillator,
 \be \psi_{i}(x)=\frac{2^{-i/2}
e^{-\frac{x^2}{2}} \mbox{H}_i(x)}{\sqrt[4]{\pi }
\sqrt{i!}}\label{baza1},\ee
 so that  the basis (\ref{baza}) is real. Hence,   and because $\textbf{H}^{\theta}$  is symmetric, then, in accordance with Section \ref{sec1},  the right and left
wavefunctions of a given resonance state can be written  in the same  form, \be
\chi^{L,R}({x}_{1},{x}_{2})=\sum_{i\geq
j}{r}_{ij}\phi_{ij}(x_{1},x_{2})=\sum_{ij}(\textbf{e})_{ij}\psi_{i}({x}_{1})\psi_{j}({x}_{2}),\label{polp}\ee with
${\sum_{i\geq j}  r_{ij}^2}=1$, where
$\{r_{ij}\}$ is the corresponding eigenvector of  $\textbf{H}^{\theta}$, and $(\textbf{e})_{ii}=r_{ii}$ and $(\textbf{e})_{ij}=2^{-1/2}r_{ij}$ $(2^{-1/2}r_{ji})$ for $i>j$ $(i<j)$. From here on we denote $\chi^{rez}=\chi^{L,R}$. As is easy to see, the reduced density matrix for particles  1 or 2
is
\be \hat{\rho}_{1 2}=\sum_{ij}(\textbf{e}^2)_{ij}|\psi_{i}\gg\ll \psi_{j} |\label{ml}.\ee  Because  the matrix $\textbf{e}$ is symmetric, its eigenvector matrix $\textbf{V}$ ($\textbf{V}^{-1}=\textbf{V}^{T})$ and eigenvalues $\textbf{D}=diag(d_{0}, d_{1},...,)$ satisfy $\textbf{e}=\textbf{V}\textbf{D} \textbf{V}^T$. Hence it is easy to infer that $\textbf{e}^2=\textbf{V}\textbf{D}^2 \textbf{V}^T$. After rewriting these formulas as
\be (\textbf{e})_{ij}=  \sum_{n}(\textbf{V})_{in}d_{n} (\textbf{V})_{jn}, \label{ttt}\ee
and
\be (\textbf{e}^2)_{ij}=  \sum_{n}(\textbf{V})_{in}d_{n}^2 (\textbf{V})_{jn}, \label{ttt1}\ee
  substituting  them into Eqs. (\ref{polp}) and (\ref{ml}), respectively, and performing some
straightforward algebra,   we   arrive at   \be
\chi^{rez}({x}_{1},{x}_{2})=\sum_{n}d_{n}u_{n}({x}_{1})u_{n}({x}_{2}),\label{jjk}\ee
and \be \rho_{12}(x,x^{'})=\ll x |\hat{\rho}_{12}| x^{'}\gg=\sum_{n}\lambda_{n}u_{n}(x)u_{n}(x^{'}),\ee
where \be u_{n}({x})=\sum_{k}
{(\textbf{{V}})_{kn}} \psi_{k}({x}),\label{jjk0}\ee
and  $\lambda_{n}=d_{n}^2$, $\ll u_{n}|u_{m}\gg=\delta_{nm}$.

The resonance parameters are determined at the angle $\theta = \theta_{opt}$
at which the eigenvalues of $\textbf{H}^{\theta}$ exhibit the most stabilized characters with
respect to $\theta$. We recall that   the resonance  energy $E^{rez}$ and  lifetime $\Gamma$  are obtained from the stable eigenvalue $W$ as $E^{rez}=\mbox{Re}[W]$ and  $\Gamma=-2\mbox{Im}[W]$, respectively. We find that the set of basis functions
 (\ref{baza}) constructed from the $90$ lowest one-particle
orbitals (\ref{baza1}) is sufficiently large  to obtain a good estimate
of the  parameters of the  lowest-energy resonance state,  at least  over the range  of $g=0$ to
$g=45$. Moreover,    in this
range the optimal value of the parameter $\theta$  is  approximately
$\theta_{opt}=0.2$, regardless of $g$.

We now examine the properties of the system. Let us first briefly discuss the special cases $g=0$ and $g\rightarrow \infty$, which correspond to the non-interacting case and the Tonks-Girardeau (TG)
 regime \cite{tg}, respectively. In these limiting situations, the     positions of the lowest resonance states are  given by $W^{g=0}=2W_{0}$
 and  $W^{TG}=W_{0}+W_{1}$, respectively,  where $W_{0}$, $W_{1}$ are   the   lowest-energy resonance positions   of the  corresponding  one-particle system,
 which we find  numerically  to be at  $W_{0}\approx0.411 - 0.0026I$, $W_{1}\approx1.014 - 0.125I$.
Here we  use the complex linear \emph{entropy} $S_{lin}$ as a measure of the correlation, $S_{lin}=1-\mbox{tr}\rho_{12}^2$.
\begin{figure}[h]
\begin{center}
\includegraphics[width=0.5\textwidth]{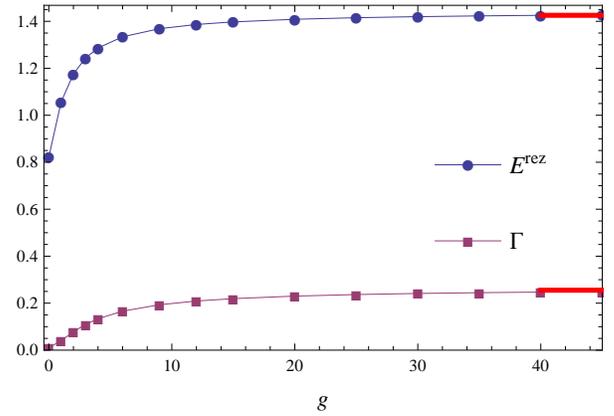}
\caption{The resonance energy and lifetime
 as  functions of $g$.
 }\label{fig:odog1}
\end{center}
\end{figure}
\begin{figure}[h]
\begin{center}
\includegraphics[width=0.5\textwidth]{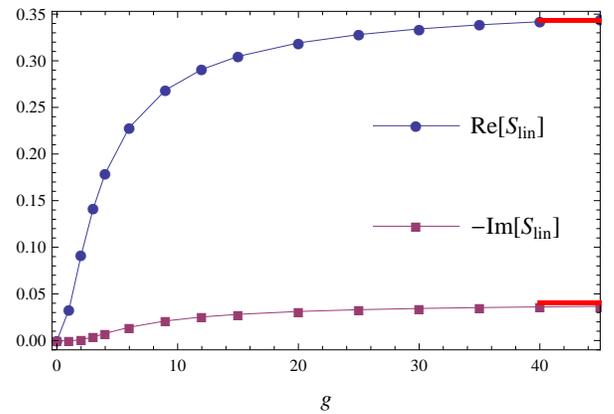}
\caption{The real and imaginary part of $S_{lin}$ as  functions of $g$. }
\label{fig:odog2}
\end{center}
\end{figure}
 Figs. \ref{fig:odog1} and \ref{fig:odog2} show our numerical results for       $W$ and for
 $S_{lin}$ as functions of $g$, respectively. The horizonal lines
 in both figures indicate the results for the TG system,  $E_{TG}^{rez}\approx1.425, \Gamma_{TG}\approx0.254$, $S_{lin}^{TG}\approx 0.34-0.04I$, where the last result  was   determined from the resonance  TG wavefunction constructed as
\begin{eqnarray}
\chi_{TG}^{rez}(x_{1},x_{2})=sgn[x_{2}-x_{1}]{1\over \sqrt{2}}det_{i=0,j=1}^{1,2}\phi_{i}^{rez}(x_{j}),
\end{eqnarray}
where  $\phi_{0}^{rez}$ and $\phi_{1}^{rez}$ ($\phi_{i}^{L,R}=\phi_{i}^{rez}$)
are   the resonance  orbitals of the one-particle  system  corresponding to $W_{0}\approx0.411-0.0026I$ and $W_{1}\approx1.014-0.125I$, respectively, $\ll\phi_{i}^{rez}|\phi_{j}^{rez}\gg=\delta_{ij}$.
In the non-interacting case, we have $E_{g=0}^{rez}\approx0.822$, $\Gamma^{g=0}\approx0.0104$ and  $\chi_{g=0}^{rez}({x}_{1},{x}_{2})=\phi_{0}^{rez}(x_{1})\phi_{0}^{rez}(x_{2})$, which gives $S_{lin}^{g=0}=0$, reflecting the
fact that there is no correlation between the particles.
We can observe  how the
results obtained for finite values of $g$  converge  to  those for the   TG system as $g$ is increased,  which, in particular, confirms the
correctness of our calculations. In fact,   the system   starts to exhibit the behaviour of the  TG system after exceeding a value of
    $g\approx40$.  We  conclude from our results
 that the larger the value of $g$, the higher the  correlations produced by two bosons, which  is  attributed to the fact that the real part of $S_{lin}$   increases  with the increase in $g$.
As  expected, the effect of changing $g$  becomes less pronounced   as   $g$ becomes larger
 and disappears in the limit  $g\rightarrow \infty$.

\section{Concluding remarks}\label{con}
We have discussed in detail the diagonal representation of a reduced density matrix determined under the framework of the CSM. Moreover, we discussed the quantification of bipartite
correlations in quantum resonance states by means of the
complex \emph{entropy}.  We also  conducted a comprehensive study of the lowest-energy resonance state of two interacting bosons trapped inside an open potential well. Among other findings,
our results show the dependence of the complex linear \emph{entropy} on the inter-boson interaction strength $g$. Its real and imaginary parts have  monotonically increasing behaviours as $g$ increases and tend to  constant values in the TG limit.

We hope our study will stimulate broader discussions of correlation in quantum resonance states.

\end{document}